\documentclass[aps,twocolumn]{revtex4}
\usepackage{amssymb}

\usepackage{graphicx}
\usepackage{graphics}
\usepackage{epsfig}

\def\b{\begin{equation}}
\def\e{\end{equation}}
\def\balll{\begin{array}{lll}}
\def\ea{\end{array}}
\def\bea{\begin{eqnarray}}
\def\eea{\end{eqnarray}}

\newcommand{\be}{\begin{equation}}
\newcommand{\ee}{\end{equation}}
\newcommand{\bqn}{\begin{eqnarray}}
\newcommand{\eqn}{\end{eqnarray}}

\begin{document}

\title{Behavior of a bipartite system in a cavity}
\author{E.R. Granhen$^{a,d}$, C.A. Linhares$^b$, A.P.C. Malbouisson$^a$
and J.M.C. Malbouisson$^c$}
\date{\today }

\begin{abstract}
We study the time evolution of a superposition of product states of
two dressed atoms in a spherical cavity in the situations of an
arbitrarily large cavity (free space) and of a small one. In the
large-cavity case, the system dissipates, whereas, for the small
cavity, the system evolves in an oscillating way and never
completely decays. We verify that the von Neumann entropy for such a
system does not depend on time, nor on the size of the cavity.
\end{abstract}

\maketitle

\affiliation{\ $^a$Centro Brasileiro de Pesquisas F\'{\i}sicas/MCT,
22290-180, Rio de Janeiro, RJ, Brazil\\$^b$Instituto de F\'{\i}sica,
Universidade do Estado do Rio de Janeiro, 20559-900, Rio de Janeiro,
RJ, Brazil\\$^c$Instituto de F\'{\i}sica, Universidade Federal da
Bahia, 40210-340, Salvador, BA, Brazil\\$^d$Faculdade de
F\'{\i}sica, Universidade Federal do Par\'{a}, 66075-110, Bel\'{e}m,
PA, Brazil}

\section{Introduction}

Stability is a main characteristic of quantum mechanical systems in
absence of interaction. When interaction with an environment is
introduced to such systems, they tend to dissipate. A material body,
for instance, an excited atom or molecule, or an excited nucleon,
changes of state in reason of its interaction with the environment.
The nature of the destabilization mechanism is in general model
dependent and approximate. An account on the subject, in particular
applied to the study of the Brownian motion, can be found for
instance in Refs.~\cite{zurek,paz}. However, stability (or not) of
quantum mechanical systems is not due only to the absence (or
presence) of interaction. For example, the behavior of atoms
confined in small cavities is completely different from the behavior
of an atom in free space or in a large cavity. In the first case,
the decay process is inhibited by the presence of boundaries, a fact
that has been pointed out since a long time ago in the literature
\cite{Morawitz,Milonni,Kleppner}, while in the second case it
completely decays after a sufficiently long elapsed time.

This phenomenon of inhibition of decay and related aspects have also
been investigated in
\cite{adolfo2,adolfo3,adolfo2a,adolfo5,linhares} using a ``dressed
state'' formalism introduced in \cite{adolfo1}. With this formalism
one recovers the experimental observation that excited states of
atoms in sufficiently small cavities are stable. In
\cite{adolfo2,adolfo3}, formulas are obtained for the probability of
an atom to remain excited for an infinitely long time, provided it
is placed in a cavity of appropriate size. For an emission frequency
in the visible red, the size of such cavity is in good agreement
with experimental observations \cite{Haroche3,Hulet}. The dressed
state formalism accounts for the fact that, for instance, a charged
physical particle is always coupled to the gauge field; in other
words, it is always ``dressed'' by a cloud of field quanta. In
general, for a system of matter particles, the idea is that the
particles are coupled to an environment, which is usually modeled in
two equivalent ways: either to represent it by a free field, as was
done in Refs.~\cite{zurek,paz}, or to consider the environment as a
reservoir composed of a large number of noninteracting harmonic
oscillators (see, for
instance,~\cite{ullersma,haake,caldeira,schramm}). In both cases,
exactly the same type of argument given above in the case of a
charged particle applies to such systems. We may speak of the
``dressing'' of the set of particles by the ensemble of the harmonic
modes of the environment. It should be noticed that our dressed
states are \textit{not} the same as those employed in optics and in
the realm of general physics usually associated to normal
coordinates \cite{bouquincohen,petrosky}. Our dressed states are
given in terms of our dressed coordinates and can be viewed as a
rigorous version of these dressing procedures, in the context of the
model employed here (see Eqs.~(\ref{qvestidas1}) and
(\ref{ortovestidas1}) in the next section).

In the present paper we study the time evolution of a two-atom
dressed state. This generalizes a previous work dealing with the
simpler situation of a superposition of states of just one atom
\cite{linhares}. Our approach to this problem makes use of the above
mentioned concept of dressed states. We will consider our system as
consisting of two atoms, each one of them interacting independently
with an environment provided by the harmonic modes of a field. The
whole system is supposed to reside in a spherical cavity of radius
$R$. We take it as a bipartite system, each subsystem consisting of
one of the dressed atoms. We will consider a superposition of two
kinds of states: either all entities (both atoms and the field
modes) are in their ground states, or just one of the atoms lies in
its first excited state, the other one and all the field modes being
in their ground states. The analysis of the density matrix of the
system leads to the time evolution of the superposed states. The
computation of the von Neumann entropy leads to the result that it
remains unchanged as the system evolves, for a cavity of any size.
We find rather contrasting behaviors for the time evolution of the
system for a very large cavity (free space, $R\rightarrow \infty $)
or for a small cavity. In the first case, as time goes on, the
system dissipates completely, while for a small cavity the departure
from the idempotency of the density matrix exhibits an oscillatory
behavior, never reaching zero.

The dressing formalism for just one atom inside a cavity is briefly
reviewed in Section 2 in order to establish basic notation and
formulas for the time evolution of the states. In Section 3, the
formalism is generalized for the two-atom system and describe the
evolution of its density matrix, either in the case of a very large
cavity (with infinite radius, that is, free space) or of a small
cavity. In Section 4, we present our conclusions.

\section{Dressing a single atom}

Let us briefly recall here some results from the analysis of
previous works for the simpler situation of just one atom, dressed
by its interaction with the environment field. We shall thus
consider an atom in the harmonic approximation, linearly coupled to
an environment modeled by the infinite set of harmonic modes of a
scalar field, inside a spherical cavity. A nonperturbative study of
the time evolution of such a system is implemented by means of
\emph{dressed} states and \emph{dressed} coordinates. We present, in
this section, a short review of this formalism, for details see
\cite{adolfo1} or \cite{livro}. We consider an atom labeled $\lambda
$, having \emph{bare} frequency $\omega _\lambda $, linearly coupled
to a field described by $N$ ($\rightarrow \infty $) oscillators,
with frequencies $ \omega _k$, $k=1,2,\ldots ,N$. The whole system
is contained in a perfectly reflecting spherical cavity of radius
$R$, the free space corresponding to the limit $R\rightarrow \infty
$. Denoting by $q_\lambda (t) $ ($p_\lambda (t)$) and $q_k(t)$
($p_k(t)$) the coordinates (momenta) associated with the atom and
the field oscillators, respectively, the Hamiltonian of the system
is taken as
\begin{equation}
H_\lambda =\frac 12\left[ p_\lambda ^2+\omega _\lambda ^2q_\lambda
^2+\sum_{k=1}^N\left( p_k^2+\omega _k^2q_k^2\right) \right] -q_\lambda
\sum_{k=1}^N\eta _\lambda \omega _kq_k,  \label{Hamiltoniana}
\end{equation}
where $\eta _\lambda $ is a constant and the limit $N\rightarrow
\infty $ will be understood later on. The Hamiltonian
(\ref{Hamiltoniana}) can be turned to principal axis by means of a
point transformation
\begin{equation}
q_{\mu }=\sum_{r_\lambda =0}^Nt_{\mu }^{r_\lambda }Q_{r_\lambda
}\,\,,\,\,\,\,\,p_{\mu }=\sum_{r_\lambda =0}^Nt_{\mu }^{r_\lambda
}P_{r_\lambda },  \label{transf}
\end{equation}
performed by an orthonormal matrix $T=(t_{\mu }^{r_\lambda })$,
where $\mu =(\lambda ,\{k\})$ , $k=1,2,\ldots ,N$, and $r_\lambda
=0,\ldots ,N$. The subscripts $\mu =\lambda $ and $\mu =k$ refer
respectively to the atom and the harmonic modes of the field and
$r_\lambda $ refers to the normal modes. In terms of normal momenta
and coordinates, the transformed Hamiltonian reads
\begin{equation}
H_\lambda =\frac 12\sum_{r_\lambda =0}^N\left( P_{r_\lambda }^2+\Omega
_{r_\lambda }^2Q_{r_\lambda }^2\right) ,  \label{diagonal}
\end{equation}
where the $\Omega _{r_\lambda }$'s are the normal frequencies
corresponding to the collective \textit{stable} oscillation modes of
the coupled system.

Using the coordinate transformation, Eq.~(\ref{transf}), in the
equations of motion derived from the Hamiltonian
Eq.~(\ref{Hamiltoniana}), and explicitly making use of the
normalization condition $\sum_{\mu =0}^N\left( t_\mu ^{r_\lambda
}\right) ^2=1,$ we get
\begin{equation}
t_k^{r_\lambda }=\frac{\eta _\lambda \omega _k}{\omega _k^2-\Omega
_{r_\lambda }^2}t_\lambda ^{r_\lambda }\;,\;\;t_\lambda ^{r_\lambda
} = \left[ 1+\sum_{k=1}^N\frac{\eta _\lambda ^2\omega _k^2}{(\omega
_k^2-\Omega _{r_\lambda }^2)^2}\right] ^{-\frac 12},  \label{tkrg1}
\end{equation}
with the condition
\begin{equation}
\omega _\lambda ^2-\Omega _{r_\lambda }^2=\sum_{k=1}^N \frac{\eta
_\lambda ^2\omega _k^2}{\omega _k^2-\Omega _{r_\lambda }^2}.
\label{Nelson1}
\end{equation}
The right-hand side of equation~(\ref{Nelson1}) diverges in the
limit $ N\rightarrow \infty $. Defining the counterterm $\delta
\omega ^2=N\eta _\lambda ^2$, it can be rewritten in the form
\begin{equation}
\omega _\lambda ^2-\delta \omega ^2-\Omega _{r_\lambda }^2 = \eta
_\lambda ^2\Omega _{r_\lambda }^2\sum_{k=1}^N\frac 1{\omega
_k^2-\Omega _{r_\lambda }^2}.  \label{Nelson2}
\end{equation}
Eq.~(\ref{Nelson2}) has $N+1$ solutions, corresponding to the $N+1$
normal collective modes. It can be shown \cite{adolfo1,livro} that
if $\omega _\lambda ^2>\delta \omega ^2$, all possible solutions for
$\Omega ^2$ are positive, physically meaning that the system
oscillates harmonically in all its modes. On the other hand, when
$\omega _\lambda ^2<\delta \omega ^2$, one of the solutions is
negative and so no stationary configuration is allowed.

Therefore, we just consider the situation in which all normal modes
are harmonic, which corresponds to the first case above, $\omega
_\lambda ^2>\delta \omega ^2$, and define the \textit{renormalized}
frequency
\begin{equation}
{\bar{\omega}}_\lambda ^2 = \lim_{N\rightarrow \infty }\left( \omega
_\lambda ^2-N\eta _\lambda ^2\right) ,  \label{omegabarra}
\end{equation}
following the pioneering work of Ref.~\cite{Thirring}. In the limit
$ N\rightarrow \infty $, equation (\ref{Nelson2}) becomes
\begin{equation}
{\bar{\omega}}_\lambda ^2-\Omega ^2 = \eta _\lambda
^2\sum_{k=1}^\infty \frac{ \Omega ^2}{\omega _k^2-\Omega ^2}.
\label{Nelson3}
\end{equation}
We see that, in this limit, the above procedure is exactly the
analogous of mass renormalization in quantum field theory: the
addition of a counterterm $ -N\eta _\lambda ^2q_\lambda ^2$
($N\rightarrow \infty $) allows one to compensate the infinity of
$\omega _\lambda ^2$ in such a way as to leave a finite, physically
meaningful, renormalized frequency ${\bar{\omega}} _\lambda $.

To proceed, we take the constant $\eta _\lambda $ as
\begin{equation}
\eta _\lambda =\sqrt{\frac{4g_\lambda \Delta \omega }\pi },
\label{eta}
\end{equation}
where $\Delta \omega $ is the interval between two neighboring field
frequencies and $g$ is the coupling constant with dimension of frequency.
The environment frequencies $\omega _k$ can be written in the form
\begin{equation}
\omega _k=k\frac{\pi c}R,\;\;\;\;k=1,2,\ldots ,  \label{discreto}
\end{equation}
and, so, $\Delta \omega =\pi c/R$. Then, using the identity
\begin{equation}
\sum_{k=1}^\infty \frac 1{k^2-u^2}=\frac 12\left[ \frac 1{u^2} -
\frac \pi u\cot \left( \pi u\right) \right] ,  \label{id4}
\end{equation}
Eq.~(\ref{Nelson3}) can be written in closed form:
\begin{equation}
\cot \left( \frac{R\Omega }c\right) =\frac \Omega {2g_\lambda } +
\frac c{R\Omega }\left( 1-\frac{R{\bar{\omega}}_\lambda
^2}{2g_\lambda c}\right) . \label{eigenfrequencies1}
\end{equation}
The elements of the transformation matrix, turning the atom--field
system to principal axis, are obtained in terms of the physically
meaningful quantities $\Omega _{r_\lambda }$ and
${\bar{\omega}}_\lambda $after some long but straighforward
manipulations~\cite{adolfo1},
\begin{eqnarray}
t_\lambda ^{r_\lambda } & = &\frac{\eta _\lambda \Omega _{r_\lambda
}}{\sqrt{ \left( \Omega _{r_\lambda }^2-{\bar{\omega}}_\lambda
^2\right) ^2+\frac{\eta _\lambda ^2}2\left( 3\Omega _{r_\lambda
}^2-{\bar{\omega}}_\lambda ^2\right)
+4g_\lambda ^2\Omega _{r_\lambda }^2}},  \nonumber \\
t_k^{r_\lambda } &=&\frac{\eta _\lambda \omega _k}{\omega _k^2-\Omega
_{r_\lambda }^2}t_\lambda ^{r_\lambda }.  \label{t0r21}
\end{eqnarray}

The eigenstates of the system atom($\lambda $)--field, $\left| l_\lambda
,l_1,l_2,...\right\rangle $, are represented by the normalized
eigenfunctions in terms of the normal coordinates $\{Q_{r_\lambda }\}$,
\begin{eqnarray}
\phi _{l_\lambda l_1l_2...}(Q,t) &=&\prod_{s_\lambda }\left[
\sqrt{\frac{ 2^{l_{s_\lambda }}}{l_{s_\lambda }!}}H_{l_{s_\lambda
}}\left( \sqrt{\frac{
\Omega _{s_\lambda }}\hbar }Q_{s_\lambda }\right) \right]  \nonumber \\
&&\times \Gamma _0^\lambda \,e^{-i\sum_{s_\lambda }\left( l_{s_\lambda
}+\frac 12\right) \Omega _{s_\lambda }t},  \label{autofuncoes}
\end{eqnarray}
where $H_{l_{s_\lambda }}$ stands for the $l_{s_\lambda }$-th Hermite
polynomial and
\[
\Gamma _0^\lambda = {\mathcal{N}}_\lambda e^{-\sum_s{}_\lambda \frac
12\Omega _{s_\lambda }Q_{s_\lambda }^2}
\]
is the normalized vacuum eigenfunction, ${\mathcal{N}}_\lambda $
being the normalization factor.

Next, \textit{dressed} coordinates $q_\lambda ^{\prime }$ and
$\{q_k^{\prime }\}$ for the \textit{dressed }atom and the
\textit{dressed} field, respectively, are introduced, defined by
\begin{equation}
\sqrt{\bar{\omega}_\mu }q_\mu ^{\prime }=\sum_{r_\lambda }t_\mu
^{r_\lambda } \sqrt{\Omega _{r_\lambda }}Q_{r_\lambda },
\label{qvestidas1}
\end{equation}
where ${\bar{\omega}}_\mu =\{{\bar{\omega}}_\lambda ,\;\omega _k\}$.
In terms of the dressed coordinates, we define for a fixed instant,
$t=0$, \textit{dressed} states, $\left| \kappa _\lambda ,\kappa
_1,\kappa _2,\cdots \right\rangle $ by means of the complete
orthonormal set of functions~\cite{adolfo1}
\begin{equation}  \label{ortovestidas1}
\psi _{\kappa _\lambda \kappa _1...}(q^{\prime })=\prod_\mu \left[
\sqrt{ \frac{2^{\kappa _\mu }}{\kappa _\mu !}}H_{\kappa _\mu }\left(
\sqrt{\frac{ \bar{\omega}_\mu }\hbar }q_\mu ^{\prime }\right)
\right] \Gamma _0^\lambda ,
\nonumber \\
\end{equation}
where, as before, $\mu $ labels collectively the dressed atom
$\lambda $ and the field modes $k=1,2,3,\ldots $, that is, $q_\mu
^{\prime }=q_\lambda ^{\prime },\,\left\{ q_k^{\prime }\right\} $.
The ground state $\Gamma _0^\lambda $ in the above equation is the
same as in Eq.~(\ref{autofuncoes} ). The invariance of the ground
state is due to our definition of dressed coordinates given by
Eq.~(\ref{qvestidas1}). Notice that the introduction of the dressed
coordinates implies, differently from the bare vacuum, the stability
of the dressed vacuum state since, by construction, it is identical
to the ground state of the interacting Hamiltonian in terms of
normal coordinates. Each function $\psi _{\kappa _\lambda \kappa
_1...}(q^{\prime })$ describes a state in which the \textit{dressed}
oscillator $q_\mu ^{\prime }$ is in its $\kappa _\mu $-th excited
state.

The particular dressed state $\left| \Gamma _1^\mu (0)\right\rangle
$ at $ t=0 $, represented by the wave function $\psi _{00\cdots
1(\mu )0\cdots }(q^{\prime })$, describes the configuration in which
\textit{only} the $\mu $-th dressed oscillator is in the
\textit{first} excited level, all others being in their ground
states. It is shown in Ref.~\cite{adolfo1}, that the time evolution
of the state $\left| \Gamma _1^\mu \right\rangle $ is given by
\begin{eqnarray}
\left| \Gamma _1^\mu (t)\right\rangle &=&\sum_\nu f_{\mu \nu }(t)\left|
\Gamma _1^\nu (0)\right\rangle \,;  \label{ortovestidas51} \\
f_{\mu \nu }(t) &=&\sum_{s_\lambda }t_\mu ^{s_\lambda }t_\nu ^{s_\lambda
}e^{-i\Omega _{s_\lambda }t},  \label{ortovestidas52}
\end{eqnarray}
with $\sum_\nu \left| f_{\mu \nu }(t)\right| ^2=1$, for all $\mu $.
This allows to interpret the coefficients $f_{\mu \nu }(t)$ as
probability amplitudes; for example, $f_{\lambda \lambda }(t)$ is
the probability amplitude that, if the dressed atom is in the first
excited state at $t=0$, it remains excited at time $t$, while
$f_{\lambda k}(t)$ represents the probability amplitude that the
$k$-th dressed harmonic mode of the field be at the first excited
level.

\section{Time evolution of a dressed two-atom state}

We now consider a bipartite system composed of two subsystems,
$\mathcal{A}$ and $\mathcal{B}$; the subsystems consist respectively
of dressed atoms $A$ and $B$, in the sense defined in the preceding
section, with $\lambda =A,B$ labeling the quantities referring to
the subsystems. The whole system is contained in a perfectly
reflecting sphere of radius $R$. In the following we consider each
atom carrying its own dressing field (a ``cloud'' of field quanta),
independently of each other. This means that we are taking the
approximation of neglecting the interaction (via the field clouds)
between them. We consider the Hilbert space spanned by the dressed
Fock-like product states,
\begin{eqnarray}  \label{Focklike}
\left| \Gamma _{n_Ak_1k_2\cdots ;\,n_Bq_1q_2\cdots }^{(AB)}\right\rangle
\equiv |n_A,k_1,k_2,\ldots ;\,n_B,q_1,q_2,\ldots \rangle  \nonumber \\
= \left| \Gamma _{n_A,k_1,k_2,\ldots }^A\right\rangle \otimes \left|
\Gamma
_{n_B,q_1,q_2,\ldots }^B\right\rangle ,  \nonumber \\
\end{eqnarray}
in which the dressed atom $A$ is at the $n_A$ excited level and the
atom $B$ is at the $n_B$ excited level; the (doubled) modes of the
field dressing the atoms $A$ and $B$ are at the $k_1,k_2,\ldots $
and $q_1,q_2,\ldots $ excited levels, respectively. Fock states of
each individual dressed atom, $A$ or $B$, possess the representation
and properties presented in the last section.

Although it is spanned by direct products of Fock states of the
parts, the Hilbert space of a bipartite system is not simply the
direct product of the Hilbert spaces of the separated parts; it
incorporates the entangled states as well. This is because quantum
mechanics relies on the assumption that a linear combination of
possible states of a given system is also an acceptable state of the
system. Therefore, many states of a bipartite system are not
separable, they cannot be reduced to an element of the direct
product of the Hilbert spaces of the separated parts; they are
entangled states which can only be conceived in a quantum mechanical
framework. We shall now concentrate in a simple family of entangled
states of the two dressed atom system.

Let us consider at time $t=0$, a family of superposed states of the
bipartite system given by
\begin{eqnarray}
\left| \Psi (0)\right\rangle & = & \sqrt{\xi }\,\left| \Gamma
_{1(A)00\cdots ;0(B)00\cdots }^{(AB)}(0)\right\rangle \nonumber \\
 && +\,\sqrt{1-\xi }\,e^{i\phi }\,\left|
\Gamma_{0(A)00\cdots ;\,1(B)00\cdots }^{(AB)}(0)\right\rangle \nonumber \\
& = & \sqrt{\xi }\,\left| 1_A,0,0,\cdots ;0_B,0,0,\cdots
\right\rangle \nonumber \\
&& +\,\sqrt{1-\xi }\,e^{i\phi }\,\left| 0_A,0,0,\cdots
;\,1_B,0,0,\cdots \right\rangle , \nonumber
\\ \label{definition-entangled-state}
\end{eqnarray}
where $0<\xi <1$. In this expression, $| \Gamma _{1(A)0(B)00\cdots
}^{(AB)}(0)\rangle $ and $| \Gamma _{0(A)1(B)00\cdots
}^{(AB)}(0)\rangle $ stand respectively for the states in which the
dressed atom $A$ ($B$) is at the first level, the dressed atom $B$
($A$) and all the field modes being in the ground state. They are
\begin{eqnarray}
\left| \Gamma _{1(A)0(B)00\cdots }^{(AB)}(0)\right\rangle & = &
\left| \Gamma _{100\cdots }^A(0)\right\rangle \otimes \left| \Gamma
_{000\cdots }^B(0)\right\rangle , \\
 \left| \Gamma
_{0(A)1(B)00\cdots }^{(AB)}(0)\right\rangle & = & \left| \Gamma
_{000\cdots }^A\right\rangle \otimes \left| \Gamma _{100\cdots
}^B(0)\right\rangle .
\end{eqnarray}
Note that, for $\xi = 1/2$ and $\phi = 0, \pi$,
states~(\ref{definition-entangled-state}) are similar to states of
the Bell basis of a bipartite system.

The two atoms are nondirectly interacting, they carry their own
dressing fields (a cloud of field quanta). The central point, which
is in the heart of the notion of entanglement, is that they share
the same common wavefunction $\left| \Psi \right\rangle $, the
superposed state. In other words, we attribute physical reality to
the superposition of the two-atom state $| \Gamma _{0(A)1(B)00\cdots
}^{(AB)}\rangle $, in which atom $B$ is at the first excited level
and the atom $A$ in the ground state, with the other state $| \Gamma
_{0(A)1(B)00\cdots }^{(AB)}\rangle $, in which the atom $A$ is at
the first excited level and the atom $B$ in the ground state;
afterwards, we study the time evolution of the system initially
described by the wavefunction
Eq.~(\ref{definition-entangled-state}). The field modes are all
taken to be in the ground state, which means that we are considering
the system at zero temperature. Since there is no interaction
between them, the atoms cannot, in both classical or
field-theoretical sense, influence one another, but as they are
described by the same wavefunction, they are in the same superposed
state and they can share \textit{information} (not mediated by field
forces). As largely stated in the literature, this is one of the more 
intriguing aspects of
quantum mechanics; the correlations predicted by the theory are not
compatible with the current idea that the state of a system, in
particular exchange of information among its subsystems, should be
mediated by interactions among them. This leads still nowadays to
different, yet controversial interpretations of quantum mechanics.

In spite of the simplicity of the model, it is widely assumed that a
pair of harmonic oscillators is a good approximation in the case of
simple atoms, for applications in quantum computing and for
experiments with trapped ions. Indeed, in the realm of quantum
computation~\cite{dajka}, a situation nearly equivalent to the one
we investigate here is studied. Two noninteracting qubits, initially
prepared in an entangled state, are \textit{coupled to their own
independent environments} and evolve under their influence. This is
quite similar to our approach, in which the time evolution of the
dressed atoms is described by Eq.~(\ref{ortovestidas51}).

At time $t$, the state of the system is described by the density
matrix $ \varrho (t)=\left| \Psi (t)\right\rangle \left\langle \Psi
(t)\right| ,$ which, using Eq.~(\ref{definition-entangled-state}),
is given by
\begin{widetext}
\begin{eqnarray}
\varrho (t)& =&\xi \left( \left| \Gamma _{100\cdots
}^A(t)\right\rangle \left\langle \Gamma _{100\cdots }^A(t)\right|
\right) \otimes \left( \left| \Gamma _{000\cdots }^B\right\rangle
\left\langle \Gamma _{000\cdots }^B\right| \right) \nonumber \\ && +
\left( 1-\xi \right) \left( \left| \Gamma _{000\cdots
}^A\right\rangle \left\langle \Gamma _{000\cdots }^A\right| \right)
\otimes \left( \left| \Gamma _{100\cdots }^B(t)\right\rangle
\left\langle \Gamma _{100\cdots
}^B(t)\right| \right)   \nonumber \\
&&+\sqrt{\xi (1-\xi )}e^{i\phi }\left( \left| \Gamma _{000\cdots
}^A\right\rangle \left\langle \Gamma _{100\cdots }^A(t)\right|
\right) \otimes \left( \left| \Gamma _{100\cdots }^B(t)\right\rangle
\left\langle \Gamma _{000\cdots }^B\right| \right) \nonumber \\
&&+\sqrt{\xi (1-\xi )}e^{-i\phi }\left( \left| \Gamma _{100\cdots
}^A(t)\right\rangle \left\langle \Gamma _{000\cdots }^A\right|
\right) \otimes \left( \left| \Gamma _{000\cdots }^B\right\rangle
\left\langle \Gamma _{100\cdots }^B(t)\right| \right) ;
\label{rho1}
\end{eqnarray}
\end{widetext}
in Eq.~(\ref{rho1}) the states $| \Gamma _{000\cdots }^A\rangle $ ,
$| \Gamma _{000\cdots }^B\rangle $ are stationary and the states $|
\Gamma _{100\cdots }^A(t)\rangle $, $| \Gamma _{100\cdots }^B(t)
\rangle \,$ evolve according to Eq.~ (\ref {ortovestidas51}).

In order to investigate how the superposed states evolve in time, we
shall consider the reduced density matrix obtained by tracing over
all the degrees of freedom associated with the field. The
computation generalizes the one presented in Ref.~\cite{linhares}.
After some long but rather straightforward calculations, we obtain
the following nonvanishing elements
\begin{eqnarray}
\rho
_{0_{\mathcal{A}}0_{\mathcal{B}}}^{0_{\mathcal{A}}0_{\mathcal{B}}}(t)
& = &1-\xi \left| f_{AA}(t)\right| ^2-(1-\xi )\left|
f_{BB}(t)\right| ^2,
\nonumber \\
\rho _{0_{\mathcal{A}}1_{\mathcal{B}}}^{0_{\mathcal{A}}1_{\mathcal{B}}}(t)
 & = &(1-\xi )\left| f_{BB}(t)\right| ^2,  \nonumber \\
\rho _{1_{\mathcal{A}}0_{\mathcal{B}}}^{1_{\mathcal{A}}0_{\mathcal{B}}}(t)
&=&\xi \left| f_{AA}(t)\right| ^2,  \label{elementos} \\
\rho _{0_{\mathcal{A}}1_{\mathcal{B}}}^{1_{\mathcal{A}}0_{\mathcal{B}}}(t)
&=&\sqrt{\xi (1-\xi )}e^{i\phi }f_{AA}^{*}(t)f_{BB}(t),  \nonumber \\
\rho _{1_{\mathcal{A}}0_{\mathcal{B}}}^{0_{\mathcal{A}}1_{\mathcal{B}}}(t)
&=&\sqrt{\xi (1-\xi )}e^{-i\phi }f_{AA}(t)f_{BB}^{*}(t).  \nonumber
\end{eqnarray}
We check immediately that the trace of this reduced density matrix is one,
\begin{equation}
\rho
_{0_{\mathcal{A}}0_{\mathcal{B}}}^{0_{\mathcal{A}}0_{\mathcal{B}
}}(t)+\rho
_{0_{\mathcal{A}}1_{\mathcal{B}}}^{0_{\mathcal{A}}1_{\mathcal{B}
}}(t)+\rho
_{1_{\mathcal{A}}0_{\mathcal{B}}}^{1_{\mathcal{A}}0_{\mathcal{B}
}}(t)+\rho
_{1_{\mathcal{A}}1_{\mathcal{B}}}^{1_{\mathcal{A}}1_{\mathcal{B}
}}(t)=1,  \label{traco11}
\end{equation}
thereby ensuring that $\rho (t)$ represents physical states of the
system. Also, we see that ${\rm Tr}\left[ \rho ^2(t)\right] \neq 1$
and, therefore, the superposed state at time $t$ is not pure. The
degree of impurity of a quantum state can be quantified by the
departure from the idempotency property. In the present case,
\begin{eqnarray}  \label{Dezao}
D(t,\xi ) &=&1-\text{Tr}\left[ \rho ^2\right]  \nonumber \\
&=&2\left( \xi \left| f_{AA}(t)\right| ^2 + (1-\xi )\left|
f_{BB}(t)\right|
^2\right)  \nonumber \\
&&-2\left( \xi \left| f_{AA}(t)\right| ^2 + (1-\xi )\left|
f_{BB}(t)\right|
^2\right) ^2.  \nonumber \\
\end{eqnarray}

In the remainder of this section we consider the two atoms as
identical and, accordingly, we adopt the subscript $0$ for both of
them, $\lambda =A=B\equiv 0$; we also take
\begin{eqnarray}
g_A=g_B\equiv g\,;\;\;\eta _A = \eta _B\equiv \eta
\,;\;\;{\bar{\omega}}_A={\
\bar{\omega}}_B\equiv \bar{\omega}  \nonumber \\
f_{AA}(t)=f_{BB}(t)\equiv f_{00}(t).
\end{eqnarray}
In this case, the matrix elements in Eqs.~(\ref{elementos}) simplify
and, from Eq.~(\ref{Dezao}), we see that the degree of impurity
becomes independent of the superposition parameter $\xi $:
\begin{equation}
D(t,\xi )=2\left| f_{00}(t)\right| ^2(1-\left| f_{00}(t)\right| ^2).
\label{Dezao1}
\end{equation}
In order to pursue the study of the time evolution of the
superposition of the two-atom states, we have to determine the
behavior of $f_{00}(t)$. We shall analyze it in the situations of a
very large cavity (free space) and of a small one.

\subsection{The limit of an arbitrarily large cavity}

We start from the matrix element $t_\mu ^{r_\lambda }$ in Eq.~(\ref
{ortovestidas51}) and consider an arbitrarily large radius $R$ for
the cavity. The two (identical) atoms behave independently from each
other, so let us focus on just one of them, either the atom $A$ or
the atom $B$, labeled $0$, so that we put $\lambda =A=B\equiv 0$.
Remembering that $\eta = \sqrt{4gc/R}$, we have
\begin{equation}
\lim_{R\rightarrow \infty }t_0^r=\lim_{R\rightarrow \infty
}\frac{\sqrt{ 4g/\pi }\Omega \sqrt{\pi c/R}}{\sqrt{\left( \Omega
^2-\bar{\omega}^2\right) ^2+4g^2\Omega ^2}}.  \label{limite-t-r-0}
\end{equation}
In this limit, $\Delta \omega = \pi c/R\rightarrow d\omega =
d\Omega$ and the sum in the definition of $f_{00}(t)$,
Eq.~(\ref{ortovestidas52}), becomes an integral, so that
\begin{equation}
f_{00}(t)=\frac{4g}\pi \int_0^\infty d\Omega \frac{\Omega
^2e^{-i\Omega t}}{ \left( \Omega ^2-\bar{\omega}^2\right)
^2+4g^2\Omega ^2}. \label{integral-f-00}
\end{equation}

Next, we define a parameter $\kappa =\sqrt{\bar{\omega}^2-g^2}$ and
consider whether $\kappa ^2> 0$ or $\kappa ^2<0$ , for which $\kappa
^2\gg 0$ and $ \kappa ^2\ll 0$ correspond respectively to weak
($g\ll \bar{\omega}_A$) and strong ($g\gg \bar{\omega}_A$) coupling
of the atoms with the environmment. For definiteness we consider in
the following $\kappa ^2> 0$, which includes the
\textit{weak-coupling} regime. We get in this case~\cite{linhares}
\begin{equation}
f_{00}(t) = e^{-gt}\left[ \cos \kappa t-\frac g\kappa \sin \kappa
t\right] +iG\left( t;\bar{\omega},g\right) ,  \label{eq27}
\end{equation}
where the function $G(t;\bar{\omega},g)$ is given by
\begin{equation}
G(t;\bar{\omega},g) = -\frac{4g}\pi \int_0^\infty dx\frac{x^2\sin
xt}{\left( x^2-\bar{\omega}^2\right) ^2+4g^2x^2}.  \label{J}
\end{equation}
For large times, the quantity $\left| f_{00}(t)\right| ^2$ is given
by~\cite{linhares}
\begin{equation}
\left| f_{00}(t)\right| ^2\approx e^{-2gt}\left[ \cos
\bar{\omega}t-\frac g{ \bar{\omega}}\sin \bar{\omega}t\right]
^2+\frac{64g^2}{\bar{\omega}^8t^6}. \label{FAA2}
\end{equation}
As $t\rightarrow \infty $, we see that the expression for $\left|
f_{00}(t)\right| ^2$ go to zero.

\subsection{Small cavity}

\begin{figure}[th]
\includegraphics[{height=5.0cm,width=7cm,angle=360}]{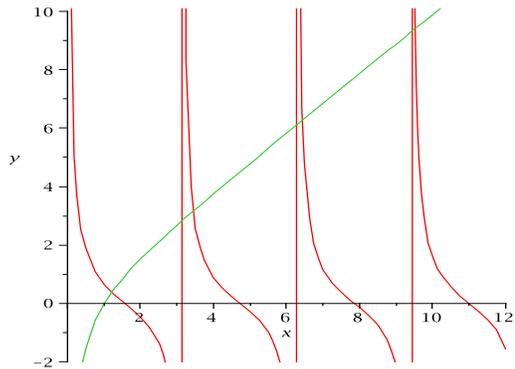}
\caption{Solutions of Eq.~(\ref{cota}), with $y=\cot (x)$ and $x=\pi
\Omega \delta /g$, for cavities satisfying the condition $\delta \ll
1$. The asymptotes of the cotangent curve correspond to the
frequencies of the field modes $\omega _k$.} \label{cotangente}
\end{figure}

For a finite (small) cavity, the spectrum of eigenfrequencies is
discrete, $ \Delta \omega $ is large, and so the approximation made
in the case of a large cavity does not apply. For a sufficiently
small cavity, the frequencies $\Omega _r$ can be determined as
follows: in Fig.~\ref {cotangente}, Eq.~(\ref{eigenfrequencies1}) is
plotted for representative values of the radius of the cavity and of
the coupling constant. We see that apart from the smallest of the
eigenfrequencies, all other ones are very close to asymptotes of the
cotangent curve, which correspond to the field frequencies. Thus let
us label the eigenfrequencies as $\Omega _0$, $\left\{ \Omega
_k\right\} $, $k=1,2,\ldots $, where $\Omega _0$ stands to the
smallest one.

Then, defining the dimensionless parameter
\begin{equation}
\delta =\frac g{\Delta \omega }=\frac{gR}{\pi c},  \label{delta}
\end{equation}
we rewrite Eq.~(\ref{eigenfrequencies1}) in the form
\begin{equation}
\cot \left( \frac{\pi \Omega \delta }g\right) =\frac \Omega {\pi g}
+ \frac g{\pi \delta \Omega }\left( 1-\frac{\delta
\bar{\omega}^2}{g^2}\right) . \label{cota}
\end{equation}
Taking $\delta \ll 1$, which corresponds to $R\ll \pi c/g$ (a small
cavity), we find that, for $k=1,2,\ldots $, the solutions are
\begin{equation}
\Omega _k\approx \frac g\delta \left( k+\frac{2\delta }{\pi k}\right) .
\label{OmegaK}
\end{equation}
If we further assume that $\Omega _0\pi \delta /g\ll 1$, a condition
compatible with $\delta \ll 1$, then $\Omega _0$ is found to be
\begin{equation}
\Omega _0\approx \bar{\omega}\left( 1-\frac{\pi \delta }3\right) .
\label{Omega0}
\end{equation}

To determine $f_{00}(t)$, we have to calculate the square of the
matrix elements $\left( t_0^0\right) ^2$ and $\left( t_k^0\right)
^2$. They are given, to first order in $\delta $, by
\begin{equation}
\left( t_0^0\right) ^2\approx \left( 1+\frac{2\pi \delta }3\right)
^{-1};\qquad \left( t_k^0\right) ^2\approx \frac 4{k^2}\frac \delta \pi
(t_0^0)^2.  \label{elem-t-k-0-aproximado}
\end{equation}
We thus obtain, for sufficiently small cavities ($\delta \ll 1$),
\begin{widetext}
\begin{eqnarray}
\left| f_{00}(t)\right| ^2 &\approx &\left( 1+\frac 23\pi \delta
\right) ^{-2}\left\{ 1+\frac{8\delta }\pi \sum_{k=1}^\infty \frac
1{k^2}\cos \left[ \bar{\omega}\left( 1-\frac{\pi \delta }3\right)
-\frac g\delta \left( k+
\frac{2\delta }{\pi k}\right) \right] t\right.   \nonumber \\
&& + \left. \frac{16\delta ^2}{\pi ^2}\sum_{k,l=1}^\infty \frac
1{k^2l^2}\cos \left[ \left( \frac g\delta -\frac{2g}{\pi kl}\right)
(k-l)\right] t\right\} .  \label{rho11}
\end{eqnarray}
\end{widetext}
To order $\delta ^2$, a lower bound for $\left| f_{00}(t)\right| ^2$
is obtained by taking the value $-1$ for both cosines in the above
formula, using the tabulated value of the Riemann zeta function
$\zeta (2)=\pi ^2/6$:
\begin{equation}
\left| f_{00}(t)\right| ^2\gtrsim \left( 1+\frac 23\pi \delta \right)
^{-2}\left\{ 1-\frac{4\pi \delta }3-\frac{4\pi ^2\delta ^2}9\right\} .
\label{rho11ss}
\end{equation}

We see, comparing Eqs.~(\ref{rho11}) and (\ref{FAA2}) that the
quantity $ \left| f_{00}(t)\right| ^2$, which dictates the behavior
of the density matrix elements and of the measure of purity in
Eq.~(\ref{Dezao1}), has very different behaviors for free space or
for a small cavity. This implies that in the situation of a small
cavity, in contrast to the free space case, all matrix elements in
Eqs.~(\ref{elementos}) are different from $zero$ at all times.

\begin{figure}[th]
\includegraphics[{height=5.0cm,width=7.0cm}]{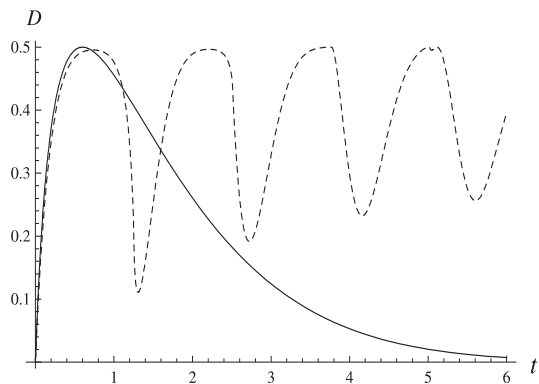}
\caption{Behavior of the degree of impurity $D$ as function of time,
equation (\ref{Dezao1}), for a small cavity (dashed line) and a very
large cavity (solid line); we take the parameters $g=0.5$, $\delta
=0.1$ and $\bar{ \omega}=1.0$ (in arbitrary units).}
\label{figDezao1}
\end{figure}

In Fig.~(\ref{figDezao1}) the degree of impurity from
Eq.~(\ref{Dezao1}) is plotted as a function of time in the cases of
an arbitrarily large cavity ($ R\rightarrow \infty $) and of a small
cavity. We take $\delta =0.1$, with $ \bar{\omega}=1.0$ and $g=0.5$
fixed (in arbitrary units).

We see from the figure that for a very large cavity (free space) the
two-atom system dissipates; with the passing of time, both atoms go
to their ground states, only the element $\rho
_{0_{\mathcal{A}}0_{\mathcal{B}}}^{0_{\mathcal{A}}0_{\mathcal{B}}}(t)
= 1$ survives in this limit. On the other hand, for a small cavity,
the system never completely decays.\\

\subsection{Time evolution of the von Neumann entropy}

We now turn our attention to the von Neumann entropy associated with
the reduced density matrix with respect to one of the subsystems; it
is obtained by taking the trace over the states of the complementary
subsystem in the full density matrix. For pure states of bipartite
systems, it measures the degree of entanglement.

The reduced density matrix for the $t=0$ superposition of states in
Eq.~(\ref {definition-entangled-state}), $\rho_{\mathcal{A}}$, is
obtained by tracing over the dressed $B$ atom. For $t\neq 0$, we
have
\begin{eqnarray}
\rho _{\mathcal{A}}(t) & = & \mathrm{Tr}_{\mathcal{B}}\left( \left|
\Psi (t)\right\rangle \left\langle \Psi (t)\right| \right) \nonumber \\
 & = & \sum_{\mu ,\nu }\xi f_{A\mu }(t)f_{A\nu }^{*}(t)\left| \Gamma
_{100\cdots }^{\mu (A)}\right\rangle \left\langle \Gamma _{100\cdots
}^{\nu (A)}\right| \nonumber \\
 & & + \,  n (1-\xi )\left| \Gamma
_{000\cdots}^A\right\rangle \left\langle \Gamma _{000\cdots
}^A\right| . \label{rhoA}
\end{eqnarray}
As time goes on, we have the time-dependent von Neumann entropy
given by
\begin{eqnarray}
E(t,\xi ) & = & -\mathrm{Tr}\left[ \rho _{\mathcal{A}}(t)\ln \rho
_{\mathcal{A} }(t)\right] \nonumber \\
 & = & -\sum_\alpha \alpha(t) \ln \alpha(t) , \label{vNt}
\end{eqnarray}
where here $\alpha(t)$ are the time-dependent eigenvalues of the
reduced density matrix. These should be solutions of the
characteristic equation, which in the case of (\ref{rhoA}), reads
\begin{widetext}
\begin{equation}
\det \left(
\begin{array}{ccccc}
1-\xi -\alpha  & 0 & 0 & 0 & \cdots  \\
0 & \xi \left| f_{AA}\right| ^2-\alpha  & \xi f_{A1}f_{AA}^{*} & \xi
f_{A2}f_{AA}^{*} & \cdots  \\
0 & \xi f_{AA}f_{A1}^{*} & \xi \left| f_{A1}\right| ^2-\alpha  & \xi
f_{A2}f_{A1}^{*} & \cdots  \\
0 & \xi f_{AA}f_{A2}^{*} & \xi f_{A1}f_{A2}^{*} & \xi \left| f_{A2}\right|
^2-\alpha  & \cdots  \\
\vdots  & \vdots  & \vdots  & \vdots  & \ddots
\end{array}
\right) =0.
\end{equation}
\end{widetext}
We thus find that the only nonzero eigenvalues of $\rho
_{\mathcal{A}}$ are
\begin{equation}
\alpha _1=1-\xi ,\qquad \alpha _2 = \xi \sum_\mu \left| f_{A\mu
}(t)\right| ^2=\xi ,
\end{equation}
which are time independent. This then implies that the von Neumann
entropy takes the expression
\begin{equation}
E(t,\xi ) = -\left[ (1-\xi )\ln (1-\xi )+\xi \ln \left( \xi \right)
\right] , \label{entropy-t}
\end{equation}
which coincides with the von Neumann entropy associated with the
initial state $\left| \Psi (0)\right\rangle $ given by
Eq.~(\ref{definition-entangled-state}); that is, all the time
dependence of the von Neumann entropy for this two-atom system,
coming from the quantities $f_{\lambda \nu }(t)$, is completely
cancelled in the computation of the entropy, in all situations, with
the maximum entanglement occuring at $\xi =1/2$. In other words,
although the superposition of states evolves in time, in very
different ways in the limits of a very large cavity and of a small
one, the entangled nature of these two-atom states remains unchanged
for all times, independently of the size of the cavity.

\section{Concluding remarks}

In this paper we have considered a system composed of two atoms in a
spherical cavity, each of them in independent interaction with an
environment field. The model employed is of a bipartite system, in
which each subsystem consists of one of the atoms dressed by its own
proper field. We make the assumption that initially we have a
superposition of two states: one in which one of the dressed atoms
is in its first excited level and the other atom and the field modes
are all in the ground state; this state is superposed with another
one in which the atoms have their roles reversed.

The time evolution of the superposition of these atomic states leads
to a time-dependent (reduced) density matrix. Expressions for its
elements are provided in both the cases of an infinitely large
cavity (that is, free space) and of a small one, when the two atoms
are considered as identical. Very different behaviors are obtained
for this time evolution. In the large-cavity case, the system shows
dissipation, and, with the passing of time, both atoms go to their
ground states. For a small cavity, an oscillating behavior is
present, so that the atoms never fully decay.

In spite of these rather contrasting behaviors and of the nontrivial
time dependence of the density matrix, we obtain a von Neumann
entropy which is independent of time and of the cavity size. We find
that the initial entanglement of the two atoms remains unchanged as
the system evolves, for a cavity of any size, in the approximation
of noninteracting dressed atoms. This could be related to the fact
that for multipartite systems the superposition principle leads
naturally to entangled states; in this case noninteracting
subsystems can thus share entangled states that hold quantum
correlations. Such quantum entanglement carries nonlocal features
which can be analyzed by comparison with classical correlations
\cite{bel1,bel2}. If an interaction between the dressed atoms,
mediated by their dressing clouds, is introduced, we expect that the
von Neumann entropy associated to the dressed atoms can depend on
time and on the size of the cavity. However, to establish the
formalism of dressed coordinates and dressed states for a system of
two interacting dressed atoms is a very hard task, which is perhaps
not possible on purely analytical grounds. We can think of
introducing this interaction as a kind of ``perturbation'' around
the individually dressed atomic states. This will be the subject of
future work.

We would like to emphasize that we here consider entanglement as a
pure quantum effect, a characteristic of quantum mechanics, which is
also nonlocal, in the sense that distant and non-interacting systems
may be entangled. This is due to the existence of superposed states,
not to the interaction between the (in our case, dressed) atoms.
Indeed such properties of entanglement of non interacting systems
have been used to conceive quantum communication
devices~\cite{bennet}.

Noninteracting systems have been, and currently are, the subject of
intense investigation in the realm of teleportation and quantum
information theory. In~\cite{lebedev}, entanglement in a mesoscopic
structure consisting of noninteracting parts is investigated. These
authors study the time-dependent electron--electron and
electron--hole correlations in a mesoscopic device and analyze the
appearance of entanglement by means of a Bell inequality test and of
Bell inequality tests based on coincidence probabilities. As we have
already mentioned before, in the framework of the theory of quantum
computating, a situation conceptually near to the one we investigate
here is studied~\cite{dajka}: two noninteracting qubits, initially
prepared in an entangled state, are \textit{coupled to their own
independent environments} and evolve under their influence. These
authors find conditions for nonvanishing entanglement at arbitrary
time, for both zero and nonzero temperatures. Also, in~Ref.
\cite{chan}, a study of the entanglement evolution of two remote
atoms interacting independently with a cavity field is presented.
In~\cite{sowa}, quantum entanglement is approached for an ensemble
of noninteracting electrons. This author uses this as a standpoint
to study the interacting gas and claims that in this context the
quantum Hall effect can be thought of as a basis for quantum
computation.

The study of entangled states of noninteracting systems is
interesting in itself. As clearly exposed in~\cite{haffner},
entanglement can exist as a purely quantum phenomenon among
\textit{noninteracting particles}, which are however described by
the same wavefunction. Entanglement means that individual particles
are not independent of each other, even if they do not interact, and
their quantum properties are inextricably ``tied up'', this being
the origin of the Schr\"{o}dinger's original denomination,
\textit{verschr\"{a}nkte Zust\"{a}nde}, for these states. In this
context, the influence of an atom on the other one is \textit{not}
due to an interaction between them, but is due to the attribution of
\textit{physical meaning to superposed states}, a concept with no
correspondence in classical physics.\\

\noindent \textbf{Acknowledgments:} The authors acknowledge CNPq/MCT
(Brazil) for partial financial support. APCM thanks FAPERJ (Brazil)
for partial financial support.

\end{document}